\documentclass[preprint,showpacs,preprintnumbers,amsmath,amssymb]{revtex4}

\usepackage{bm}
\usepackage[colorlinks=true,linkcolor=blue,citecolor=blue,urlcolor=blue]{hyperref}
\usepackage{color}

\usepackage{graphicx}
\usepackage{epstopdf}

\begin{document}
	

	\title{Preparation of high crystalline nanoparticles of rare-earth based complex  pervoskites and comparison of their structural and magnetic properties with bulk counterparts}
	
	
	\author{M. A. Basith}
	\email[Author to whom correspondence should be addressed (e-mail): ]{mabasith@phy.buet.ac.bd}
	\author{M. A. Islam}
	\affiliation{Department of Physics, Bangladesh University of Engineering and Technology, Dhaka-1000, Bangladesh.
	}
	\author{Bashir Ahmmad}
	\affiliation{Graduate School of Science and Engineering, Yamagata University, 4-3-16 Jonan, Yonezawa 992-8510, Japan.}
	
	\author{Md. Sarowar Hossain }
	\affiliation{S. N. Bose National Centre for Basic Sciences, Salt Lake City, Kolkata, West Bengal 700098, India.
	}

	\author{K. M{\o}lhave}
	\affiliation{Department of Micro-and Nanotechnology, Technical University of Denmark, Kgs. Lyngby 2800, Denmark.}


	\date{\today}

\begin{abstract}
A simple route to prepare  Gd$_{0.7}$Sr$_{0.3}$MnO$_3$ nanoparticles by ultrasonication of their bulk powder materials is presented in this article. For comparison, Gd$_{0.7}$Sr$_{0.3}$MnO$_3$ nanoparticles are also prepared by ball milling. The prepared samples are characterized by X-ray diffraction (XRD),
field emission scanning electron microscope (FESEM), energy dispersive X-ray (EDX), X-ray photoelectron spectroscope (XPS), and Superconducting Quantum Interference Device (SQUID) magnetometer. XRD Rietveld analysis is carried out extensively for the determination of crystallographic parameters and the amount of crystalline and amorphous phases. FESEM images demonstrate the formation of nanoparticles  with average particle size in the range of 50-100 nm for both ultrasonication and 4 hours (h) of ball milling. The bulk materials and nanoparticles synthesized by both ultrasonication and 4 h ball milling exhibit a paramagnetic to spin-glass transition. However, nanoparticles synthesized by 8 h and 12 h ball milling do not reveal any phase transition, rather show an upturn of magnetization at low temperature. The degradation of the magnetic properties in ball milled nanoparticles may be associated with amorphization of the nanoparticles due to ball milling particularly for milling time exceeding 8 h. This investigation demonstrates the potential of ultrasonication as a simple route to prepare high crystalline rare-earth based manganite nanoparticles with improved control compared to the traditional ball milling technique.

\end{abstract}

\maketitle
\section{Introduction} \label{I}

Magnetic properties of mixed valence perovskite manganites R$_{1-x}$A$_{x}$MnO$_3$ (R = La, Gd, Pr, Nd, Sm etc and A = Sr, Ba and Ca) at the nanometer scale are currently the focus of intense investigations due to their interesting physical properties as well as potential technological applications \cite{ref51,ref3, ref501}. Along with a wide range of functional properties like colossal magnetoresistance, magnetocaloric effect, the perovskite class of manganites also exhibit multiferroic \cite{ref17, ref79} properties which is technologically very important. Therefore, the past decade has seen an increased interest in the study of multiferroic  perovskites \cite{ref17,ref19}. A spontaneous electric polarization in the presence of magnetic fields was reported in Gd$_{1-x}$Sr$_{x}$MnO$_3$ perovskite manganites \cite{ref20}. When the size of magnetic particles is reduced to a few tens of nanometers, they exhibit fascinating magnetic and electronic properties that are significantly different from their bulk counterparts \cite{ref51,ref5,ref88, ref78}. In these manganite systems, novel magnetic properties such as colossal magnetoresistance \cite{ref52} can be obtained around the ferromagnetic to paramagnetic transition temperature (T$_c$) \cite{ref53}. However, the transition temperatures, i.e. the particle size dependent T$_c$ values for the same material system were found to vary in different investigations \cite{ref53,ref55,ref10}. The conflicting results might be due to the influence of the fabrication process as well as different oxygen content of manganites \cite{ref51}. It is reported that the T$_c$ values of the samples can be decreased by adopting different fabrication techniques for the preparation of manganite nanoparticles due to very small variations in the chemical composition and oxygen stoichiometry \cite{ref3}. It is also reported that in magnetic materials, the magnetic order in the surface of the particles is modified by forming a magnetically dead layer \cite{ref3, ref91, ref98}, and consequently affect the magnetization, transition temperature and anisotropy of the material system \cite{ref3}. By using the available wet chemical methods, it is difficult to control the complex solution processes and  the size of the synthesized nanoparticles \cite{ref9}. Therefore, the synthesis of manganite nanoparticles with controlled size, chemical composition, oxygen stoichiometry  and unmodified surface is of fundamental and technological interests. 

The perovskite manganite nanoparticles of various compositions have been synthesized using the ball milling technique \cite{ref10,ref11, ref60, ref61, ref81} to produce nanoparticles from a few to tens nanometers \cite{ref10}. However, the magnetic properties of the finer particles were found to degrade.  The degradation was ascribed to the increase of the defect density and surface roughness of La$_{0.8}$Sr$_{0.2}$MnO$_{3-\delta}$ nanoparticles due to the high energy ball milling \cite{ref11}. During the fabrication of La$_{0.67}$Sr$_{0.33}$MnO$_{3}$ manganite particles using ball milling, for a milling time longer than 40 h, the perovskite structure disappeared and a completely amorphous phase was formed \cite{ref60}. Moreover, the magnetic measurements showed that ball milled samples had an inhomogeneous magnetic state \cite{ref60}.	

Notably, the utilization of ultrasonic energy in a process called sonofragmentation offers a facile, versatile synthetic tool for the preparation of nanostructured materials that is often difficult by conventional methods \cite{ref13,ref701, ref702, ref703}.  Ultrasonic dispersion has extensive use to disperse sub-micron agglomerated powders in liquid suspensions. It is possible to  break down the aggregates of nanocrystalline particles  by using the effects of ultrasound which  generates many localized hot spots with the particles within the solution \cite{ref704} and during the process, the implosive collapse of the bubbles causes an inward rush of liquid known as ‘shockwaves and microsteaming’ in which high velocity is produced. Thus the agglomerates can be broken down utilizing the effects of ultrasound \cite{ref705}, resulting in reduction of particle size. Recently, we have synthesized Bi based Bi$_{0.9}$Gd$_{0.1}$Fe$_{1-x}$Ti$_{x}$O$_3$  nanoparticles by ultrasonication of their micro-meter sized multiferroic bulk powder materials \cite{ref13} with particle size that can be varied as a function of sonication time. Transmission electron microscopy imaging confirmed the formation of ultrasonically prepared single crystalline nanoparticles with a mean size of 11-13 nm for 60 minutes sonication.

In the present investigation, we have synthesized nanoparticles of rare-earth based  Gd$_{0.7}$Sr$_{0.3}$MnO$_3$ by ultrasonication of their bulk powder materials. For comparison, we have also synthesized Gd$_{0.7}$Sr$_{0.3}$MnO$_3$ nanoparticles by planetary ball milling. This compound was chosen due to the fact that Gd$_{0.7}$Sr$_{0.3}$MnO$_3$ contain significant amount of heavy rare earth Gd and the magnetic properties of rare-earth based manganites are quite different from that of Bi based materials as reported elsewhere \cite{ref59, ref92}. The magnetic properties of Bi based manganites are attributed to the presence of highly polarizable 6s${^2}$ lone  pair of electrons present on the Bi atom, which strongly decreases the  mobility of e$_g$ electrons \cite{ref59, ref92}. Whereas the magnetic properties of rare earth based mixed valence perovskite manganites involve simultaneous transfer of an itinerant e$_g$ electron from the Mn$^{3+}$ to the oxygen and from the oxygen to the neighbouring Mn$^{4+}$  \cite{ref94}. Moreover, in these manganite systems the conversion of Mn$^{3+}$ to Mn$^{4+}$ is significantly influenced by oxygen vacancies \cite{ref93}.  Till now, the perovskite Gd-Sr manganites \cite{ref56} have been well studied in the bulk form as a single crystal \cite{ref57} and polycrystalline samples \cite{ref52}. However, due to the limitation of preparation techniques that allow reproducible fabrication of high crystalline nanoparticles of complex rare-earth based pervoskite Gd$_{0.7}$Sr$_{0.3}$MnO$_3$ manganites, many of the novel properties of these nanostructures are yet to be unveiled.  Therefore, in this investigation, rare-earth based Gd$_{0.7}$Sr$_{0.3}$MnO$_3$ nanoparticles were produced by ultrasonication as well as conventional ball milling techniques. The structural and magnetic properties of the synthesized nanoparticles were investigated and compared to their corresponding bulk materials.

\section{Experimental details} \label{II}

The perovskite manganite with nominal composition  Gd$_{0.7}$Sr$_{0.3}$MnO$_3$  was prepared initially by conventional solid state reaction technique which was described in details elsewhere \cite{ref14,ref15}. The analytical grade Gd$_{2}$O$_3$, SrCO$_3$ and MnCO$_3$ powders were mixed and ground in an agate mortar till a homogeneous mixture was formed. This mixture was calcined at around 1100 $^{0}$C for 18 hours in a programmable furnace with intermediate grinding after 12 hours. The powders were pressed into pellets of thickness 1 mm and diameter 10 mm by using a hydraulic press and sintered at 1300 $^{0}$C for 6 hours \cite{ref14}. The sintered powder and pellets were used to measure the required structural, morphological and magnetic properties of the bulk materials. In order to prepare the nanoparticles by using ultrasonication and ball milling techniques, the ceramic pellets were ground again into powder by manual grinding to obtain powder materials. 

\begin{figure}[!h]
	\centering
	\includegraphics[width=16cm]{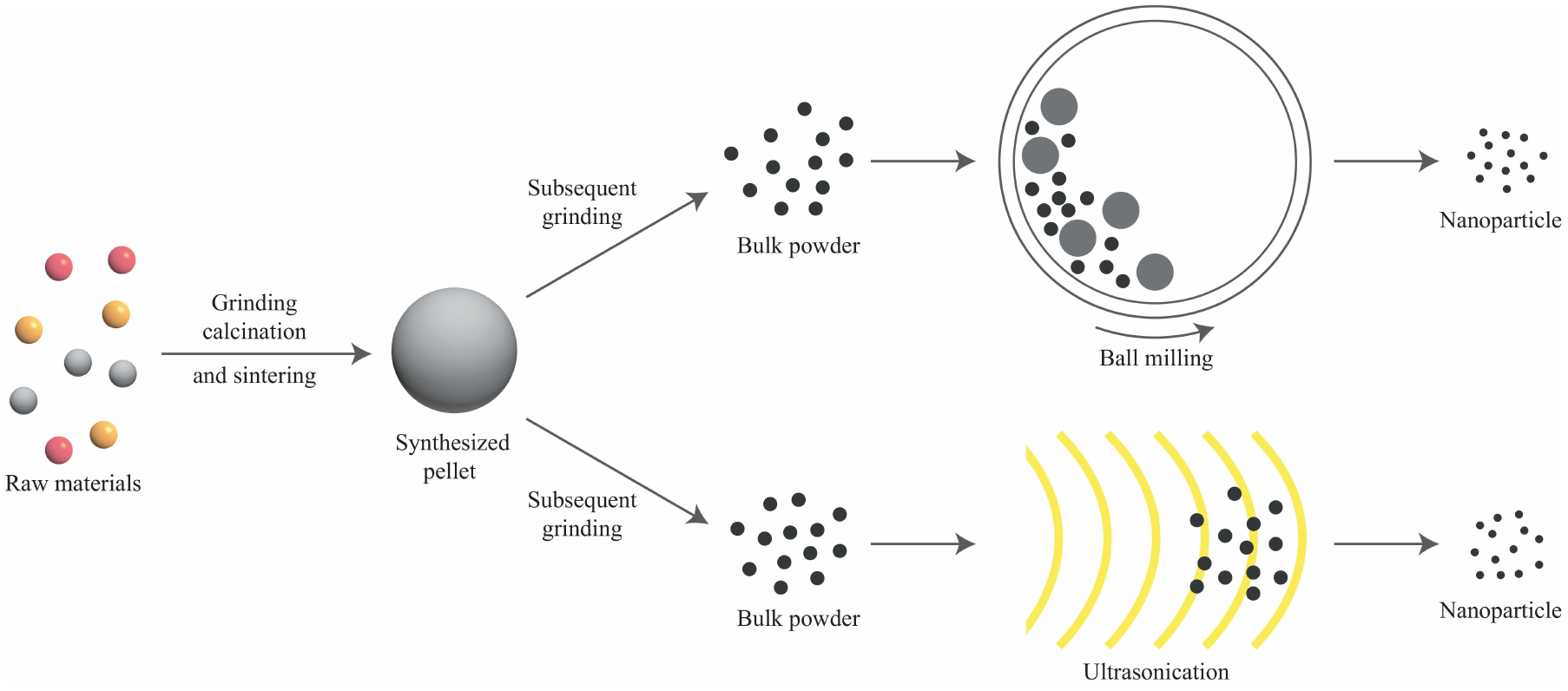}
	\caption{A scheme delineating the preparation of Gd$_{0.7}$Sr$_{0.3}$MnO$_3$ nanoparticles by using (a) ultrasonication and (b) ball milling techniques.} \label{Fig:block}
\end{figure}

A portion of the micro-meter size powder was subsequently mixed with isopropanol. The mixtures of isopropanol and powder with a mass percentage of $\sim$ 0.5 \%  were put into an ultrasonic bath (power 50 W) and sonicated for 60 minutes. After six hours rest, $\sim$ 30 \% of the mass was collected as supernatant from the mixture and dried naturally for the required characterization. Another portion of the powder materials was used to prepare nanoparticles of Gd$_{0.7}$Sr$_{0.3}$MnO$_3$ manganites by using ball milling technique as was described in the published reports \cite{ref10, ref11}. The bulk polycrystalline Gd$_{0.7}$Sr$_{0.3}$MnO$_3$ powder was ball milled for 4-12 hours by conventional planetary ball mill (MTI corporation, Model: QM-3SP2) with stainless steel vials and zirconia balls. The ratio of the ball and powder weight was 50:1 and the rotating speed was set to 300 rpm. The preparation of Gd$_{0.7}$Sr$_{0.3}$MnO$_3$ nanoparticles by using ultrasonication and ball milling techniques is illustrated schematically in figure~\ref{Fig:block}.

The crystal structures of the bulk powder materials, nanoparticles prepared by ultrasonication and also prepared by ball milling were determined from X-ray diffraction (XRD) data. XRD patterns were collected at room temperature using a diffractometer (Rigaku SmartLab) with CuK$_{\alpha}$  (${\lambda}$ = 1.5418 $\AA$) radiation. Rietveld analysis of XRD data was carried out using the ‘FULLPROF’ package \cite{ref101}. The microstructure of the surface of the sintered pellets, and the size and distribution of the synthesized nanoparticles were observed using a field emission scanning electron microscope (FESEM, JEOL, JSM 5800). X-ray photoelectron spectroscopy (XPS, ULVAC-PHI Inc., Model 1600) analysis was carried out with a Mg-K{$\alpha$} radiation source. The temperature dependent magnetization measurements of Gd$_{0.7}$Sr$_{0.3}$MnO$_3$ bulk materials and nanoparticles were carried out using a Superconducting Quantum Interference Device (SQUID) magnetometer (Quantum Design MPMS-XL7, USA) both at zero field cooling (ZFC) and field cooling (FC) processes. The field dependent magnetization measurements  were also carried out using the same SQUID magnetometer at 20 K and 300 K.

\section{Results and discussions} \label{III}
\subsection{Structural Analysis}

\begin{figure}[!h]
	\centering
	\includegraphics[width=12cm]{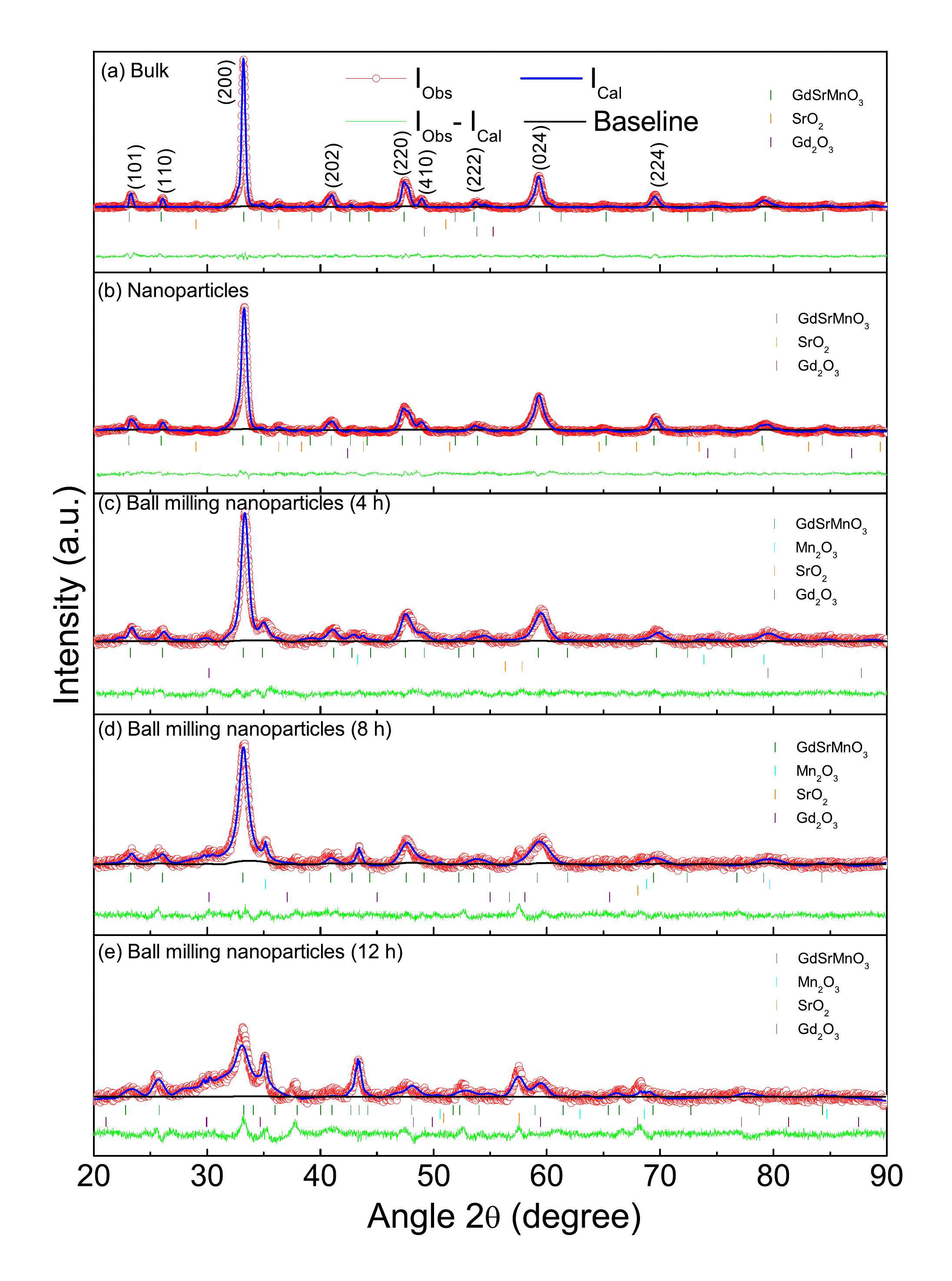}
	\caption{Rietveld plots of XRD patterns of Gd$_{0.7}$Sr$_{0.3}$MnO$_3$ (a) bulk materials (b) nanoparticles prepared by ultrasonication and ball milling of bulk powder materials for (c) 4 h, (d) 8 h and (e)12 h. In (a-e) the circles are the observed profile, (I$_{Obs}$), the solid blue line is the calculated pattern (I$_{Cal}$) and the bottom curve is the difference between the observed and calculated patterns (I$_{Obs}$-I$_{Cal}$). In (a-e) the small vertical marks below each profile show the Bragg position.} \label{fig1}
\end{figure}

The measured and calculated XRD intensity patterns after Rietveld refinement of  Gd$_{0.7}$Sr$_{0.3}$MnO$_3$ bulk materials, and nanoparticles prepared by ultrasonication and ball milling techniques for 4 h, 8 h and 12 h milling time are shown in figures \ref{fig1} (a-e) respectively. The Bragg positions of the reflections of Gd$_{0.7}$Sr$_{0.3}$MnO$_3$ and other oxides are indicated by vertical lines below the pattern. The lattice parameters and cell volumes of bulk materials and nanoparticles prepared by ultrasonication and ball milling techniques are presented in Table \ref{Tab1}. The structure of Gd$_{0.7}$Sr$_{0.3}$MnO$_3$ manganites is an orthorhombically distorted perovskite structure (spatial group Pbnm) determined by Rietveld refinement and is in agreement with the structure observed in related compounds \cite{ref52,ref56}. The lattice parameters for the bulk polycrystalline sample are: a = 5.399 (4) ${\buildrel _{\circ} \over {\mathrm{A}}}$, b = 5.472 (1) ${\buildrel _{\circ} \over {\mathrm{A}}}$ and c = 7.625 (1) ${\buildrel _{\circ} \over {\mathrm{A}}}$. The lattice parameters are:  5.399 (1)  ${\buildrel _{\circ} \over {\mathrm{A}}}$, 5.478 (1) ${\buildrel _{\circ} \over {\mathrm{A}}}$ and 7.607 (1) ${\buildrel _{\circ} \over {\mathrm{A}}}$ for the corresponding nanoparticles prepared by ultrasonication. The lattice parameters are in good agreement with the reported parameters of orthorhombic Gd$_{0.5}$Sr$_{0.5}$MnO$_3$ \cite{ref52} manganites. In order to quantitatively express the microscopic distortion relative to the ideal perovskite structure, 
orthorhombic deformation (D \%) defined as
\begin{equation}
D=\frac{1}{3}\sum_{i=1}^{3}|\frac{a_i-\overline{a}}{a_i}|\times 100\\
\end{equation}
where a$_1$=a, a$_2$=b, a$_3$=c/$\sqrt{2}$, and $\overline{a}$=(abc/$\sqrt{2}$)$^{1/3}$ \cite{ref63}, have also been calculated and inserted in Table \ref{Tab1}. The D \% was nominally changed for ultrasonically prepared and 4 h ball milled nanoparticles. However, for milling time higher than 8 h, D \% has increased significantly.

\begin{table}[!h]
	\caption{Structural parameters (obtained from rietveld refinement; $\chi ^2$ indicates goodness of fitting) and Orthorhombic deformation D(\%) of the system Gd$_{0.7}$Sr$_{0.3}$MnO$_3$ containing orthorhombic phase symmetry.} \label{Tab1} 
	\begin{center}
		\begin{tabular}{l l l l l l l l l }
			\hline

			Samples & a$_{orth}$ $\mathrm{(\AA)}$ & b$_{orth}$ $\mathrm{(\AA)}$  & c$_{orth}$ $\mathrm{(\AA)}$ & Unit cell & Orthorhombic & Fitting  \\
			&  &   & & Volume $\mathrm{(\AA ^3)}$  & Deformation & Parameters \\
			&  &      &   & &  D (\%)  &  \\
			\hline
			
			& &	 &	 &	 &	 & $\chi ^2$=6.95	\\
			Bulk &5.399 (4) &	5.472 (1) &	7.625 (1) &	225.3 (1) &		0.63 & R$_B$=6.62	\\
			&	 &	 &	 &	 &		 &	R$_f$=3.81\\
			\\

			& &	 &	 &	 &	 	 & $\chi ^2$=3.54	\\
			Nano &5.399 (1) &	5.478 (1)&	7.607 (1)&	225 (1) &			0.73 &		 R$_B$=3.62\\
			&	 &	 &	 &	 &		 &	R$_f$=3.25\\

			\\
			
			& &	 &	 &	 &	 	 & $\chi ^2$=5.62	\\
			4h &5.382 (2) &	5.462 (2) &	7.589 (3) &	223.1 (1) &		0.72 & R$_B$=6.90\\
			&	 &	 &	 &	 &	 	 &	R$_f$=3.76\\
			
			\\
			
			& &	 &	 &	 &	 	 & $\chi ^2$=5.42	\\
			8h &5.373 (2) &	5.472 (2) &                        	7.607 (4) &	223.6 (1) &		0.79 & R$_B$=3.67	\\
			&	 &	 &	 &	 &	 	 &	R$_f$=2.52\\
			
			\\
			
			& &	 &	 &	 &	 	 & $\chi ^2$=6.53	\\
			12h &5.393 (3) &	5.616 (2) &	7.565 (4) &	229.6 (2) &			1.99 &	R$_B$=2.93 \\
			&	 &	 &	 &	 &	 	 &	R$_f$=9.24\\

			\hline
			\hline
		\end{tabular}
	\end{center}
\end{table}

\begin{table}[!h]
\caption{Wt\% of the phases present in Gd$_{0.7}$Sr$_{0.3}$MnO$_3$ samples estimated from the X'Pert HighScore Plus and Rietveld analysis. The table also shows various R factors such as R$_{p}$ (profile factor), R$_{wp}$ (weighted profile factor) and Gof (goodness-of-fit)}. \label{Tab2} 
	\begin{center}
		\begin{tabular}{l l l l l l l l l l l l}
			\hline
			
			Phases &	Bulk (100\%)  &	Nano (100\%)  & 4h (100\%)  & 8h (100\%)  & 12h (100\%)  \\
			
			\hline	
			FWHM &	0.383 &	0.532 &	0.775 &	0.905&	0.964\\
		Microstrain \% &	0.0242 (4) &	0.0408 (7) &	0.0833 (1) &	0.0975 (4)&	0.1176 (2)\\
			
			Amorphousity &12.4 (6) &	13.6 (1) &	14.9 (6) &	33.2(5) &	35.9 (9) \\

			Wt\%  GdSrMnO$_3$
			Orthorombic (pbnm) &84.5(6)&	83.2(5) &	78.1(8)&	61.7 (4)&	61.9 (4)\\

			Wt\% Mn$_2$O$_3$
			Cubic (la3) & 0&	0&	5.2(8)&	1.0&	0.3 \\

			Wt\% SrO$_2$
			Tetragonal (l4/mmm) &3.0 &	2.9&	1.4(1) &	0.1&	0.1\\

			Wt\% Gd$_2$O$_3$
			Cubic (la-3) &	0.1&	0.1&	0.4&	4(2)&	 1.8 (1) \\
			
			R$_p$ (\%) &18.2&	17.3&	22.9&	21.0& 17.5\\
			
			R$_{wp}$ (\%)	& 11.6 &	10.8&	16.3&	17.0&	15.2\\
			
			
			Gof (\%) &2.63&	1.88&	2.37&	2.32&	2.55\\
			
			\hline
			\hline
		\end{tabular}
	\end{center}
\end{table}

XRD analysis is one of the common ways to investigate the nature of crystalline and amorphous phases of a material \cite{ref71}. The peaks intensity of the ultrasonically prepared nanoparticles slightly decrease together with a little broadening of the peaks, which indicate a decrease in crystallinity of the nanoparticles. On the contrary, the intensity peaks of the ball milled particles decrease drastically with milling time  along with a large broadening of the peaks as shown in figures \ref{fig1} (c-e). This indicates a notable reduction in the crystallinity of the materials due to ball milling  \cite{ref210}. The width of the (200) peak of the bulk materials is 0.383, which was increased to 0.532 for ultrasonically prepared nanoparticles. The width of the same peaks for ball milling nanoparticles gradually increase with prolongation of the milling time. Table \ref{Tab2} shows the values of the FWHM of (200) peaks for all materials under scrutiny. While performing this calculation, instrumental peak broadening factor has been taken into consideration. To determine the instrumental broadening, a diffraction profile from a standard material like silicon has been collected as a reference. In previous investigations \cite{ref601,ref602}, the instrumental broadening was corrected, corresponding to the peaks of the sample using the relation:
\begin{equation}
\beta = \sqrt{\beta_{exp}^2 - \beta_{inst}^2}
\end{equation}
where, $\beta$, $\beta_{exp}$ and $\beta_{inst}$ are the FWHMs of the intrinsic profile, the experimental profile and the instrumental broadening profile respectively at the same value of $2 \theta$. In fact, the FWHMs of the obtained XRD peaks from the XRD machine are not simply the summation of the actual FWHM and FWHM of the reference. Rather the obtained response is the convoluted result of the desired intrinsic sample profile along with the instrumental broadening profile \cite{ref603, ref604}. Therefore, instead of simplified subtraction, using deconvolution operation, the effect of instrumental peak broadening has been minimized, and corrected FWHM of the (200) peaks of the samples were inserted in Table \ref{Tab2}. By using a block diagram the deconvolution operation system has been described in supplemental information \cite{ref99}. It is noteworthy that lattice strain, which is a measure of the distribution of lattice constants arising from crystal imperfections and distortions, also affect the broadening of Bragg peak. Therefore, the microstrain of the synthesized samples was calculated using Williamson-Hall analysis \cite{ref904} and the values are inserted in Table \ref{Tab2}. Like orthorhombic deformation, the micro-strain was also found to increase gradually with milling time \cite{ref605} and for 12 h ball milling sample the microstrain is significantly higher. 




The broad and diffuse peaks particularly centered at 33$^{0}$ figures \ref{fig1} (b)-(e)) is a clear indication of crystalline to amorphous like phase transition of the bulk powder materials after sonication as well as ball milling. In a previous investigation, the intensity of peaks belonging to La$_{0.67}$Sr$_{0.33}$MnO$_{3}$ manganites was also found to decrease gradually for a milling time $>$ 12 hours \cite{ref60}. The decrease in intensity of peaks suggested the appearance of an amorphous phase in La$_{0.67}$Sr$_{0.33}$MnO$_{3}$ manganites when the milling time is longer than 12 hours \cite{ref60}. Notably, although XRD patterns provide indication of the formation of crystalline to amorphous like phase, however, the quantification of the amount of crystalline and amorphous phases is  difficult \cite{ref64}. In this investigation, using the Rietveld refinement procedure, the quantity of amorphous and crystalline phases of the synthesized samples have been calculated. Amorphousity has been calculated by the intensity ratio of the diffraction peaks and sum of all measured intensity with the help of X'Pert HighScore Plus where the program has been calibrated to the crystallinity analysis for a standard sample. But a completely crystalline sample has some background intensity, which arises from imperfections of the sample, the X-ray optics of the instrument, sample fluorescence and scattering. Therefore, X'Pert HighScore Plus, the background is calibrated with the granularity of 22 and a bending factor of 5 hence is applied to the correction for all the samples. The following equation was involved for amorphousity calculation
\begin{equation}
Amorphousity \% = 100-[100 \times \Sigma  I_{net.} / (\Sigma I_{tot} - I_{const.bgr})]
\end{equation}
where $I_{net.}$: crystal intensity; $I_{const.bgr}$: background intensity; $I_{tot}$: total intensity.  Then the wt\% of crystalline and amorphous phases have been normalized (100\%) to obtain the wt\% of all other phases inside the samples. The wt\% of crystalline and amorphous phases are inserted in Table \ref{Tab2}. The amorphization \% nominally increase for nanoparticles prepared by ultrasonication and 4 h ball milling compared to that of bulk materials. However, for a milling time of 8 h and 12 h, the amorphous phase formation is nearly three times higher compared to that of bulk materials. For 8 h and 12 h milling time, the amorphous phase is around 33.2 wt$\%$ and  35.9 wt$\%$, respectively as shown in Table \ref{Tab2}.




Rietveld refinement has also been performed to estimate the amount of other oxides presence in these manganites. The wt$\%$ of other oxides obtained after refinement was inserted in Table \ref{Tab2}. In the bulk materials apart from orthorhombic Gd$_{0.7}$Sr$_{0.3}$MnO$_3$ (84.5 wt$\%$) manganites, the refined data shows the presence of SrO$_2$ (3 wt$\%$). For 4 h ball milled nanoparticles, along with orthorhombic Gd$_{0.7}$Sr$_{0.3}$MnO$_3$ (78.1 wt$\%$) manganites, a fair amount of Mn$_2$O$_3$ (5.2 wt$\%$) is formed. For a longer milling time of 8 h and 12 h, the amount of Mn$_2$O$_3$ is reduced significantly, however, the amorphization, orthorhombic deformation and microstrain  increase noticeably as mentioned earlier. In the ultrasonically prepared nanoparticles, the orthorhombic Gd$_{0.7}$Sr$_{0.3}$MnO$_3$ phase is 83.2 wt$\%$ along with a small amount of other oxides. In figures \ref{fig1} (a) - (e), the vertical lines demonstrate the Bragg positions of these oxide materials along with orthorhombic Gd$_{0.7}$Sr$_{0.3}$MnO$_3$ manganites. 


\begin{figure}[!h]
	\centering
	\includegraphics[width=16cm]{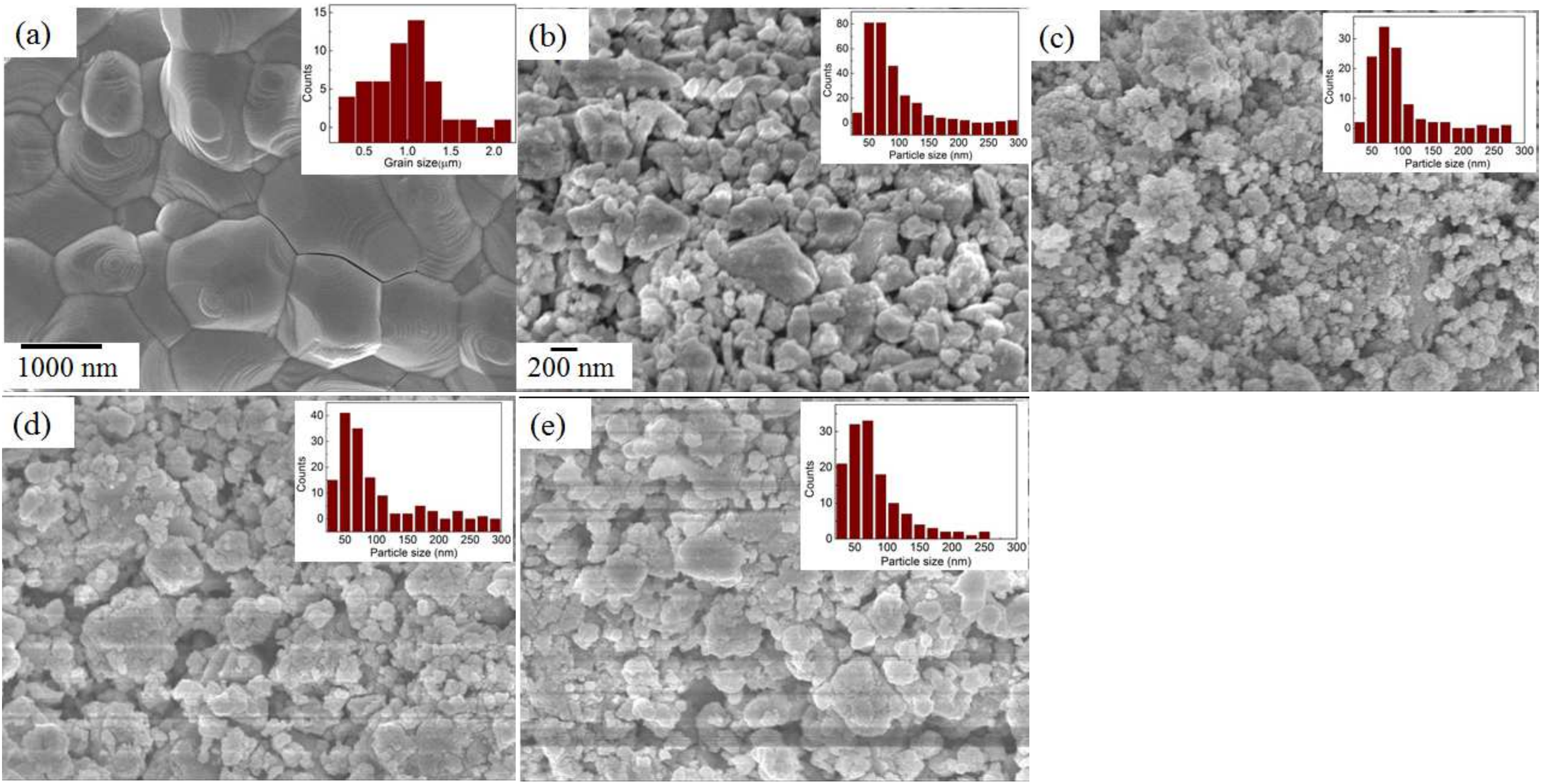}
	\caption{(a) FESEM image of the surface morphology of bulk polycrystalline Gd$_{0.7}$Sr$_{0.3}$MnO$_3$ manganites. (b) FESEM image and particle size distribution histogram of Gd$_{0.7}$Sr$_{0.3}$MnO$_3$ nanoparticles prepared by ultrasonication \cite{ref13}. (c-e) FESEM images and particle size histograms of Gd$_{0.7}$Sr$_{0.3}$MnO$_3$ nanoparticles prepared by 4 h, 8 h and 12 h ball milling, respectively.} \label{fig2}
\end{figure}
Figures \ref{fig2}(a) demonstrates the surface morphology of the pellets of Gd$_{0.7}$Sr$_{0.3}$MnO$_3$ bulk perovskite manganites. The surface of the pellets are pretty homogeneous and the average grain size is around 1-1.5 $\mu m$. Figure \ref{fig2}(b) shows FESEM image of Gd$_{0.7}$Sr$_{0.3}$MnO$_3$ particles prepared by ultrasonication, images (b-e) show particle size for 4 h, 8 h and 12 h ball milling, respectively. The distribution of the ultrasonically and ball milled nanoparticles deduced from micrographs are also shown in images (b-e). The size of the ultrasonically prepared nanoparticles is estimated from the distribution histogram of figure \ref{fig2}(b) is 50-100 nm. The average size of the nanoparticles prepared by 4 h ball milling is also ranging from 50-100 nm. For a further increase of the milling time, the overall distribution of the particle size histogram did not change significantly, however, agglomeration of powder materials increase with milling time \cite{ref211}. In the present investigation, the calculated size of the ultrasonically prepared particles from the FWHM of (200) peak of XRD pattern using Scherrer's formula \cite{ref16} is $\sim {~}$27 nm which is much smaller than the value observed from FESEM image. The large particle size determined by electron microscopy images compared to that calculated by Scherrer equation has also been reported in previous investigations \cite{ref301, ref302} and this was due to the agglomeration of the particles. For ball milled nanoparticles, broadening of the (200) peak is also associated with microstrain as well as amorphization of the material. Therefore, we have not calculated the crystallite size of the ball milled nanoparticles using Scherrer equation. 

Notably, the EDX analysis presented also in the supplemental Table 1 \cite{ref99} reveals the presence of Gd, Sr, Mn, and O elements in bulk materials as well as in ultrasonically prepared nanoparticles. However, the compositional analysis of the ball milled nanoparticles also confirms the presence of Zr along with Gd, Sr, Mn, and O. The Zr was contaminated during ball milling and the amount of Zr was increased with increasing milling time. 
	
\begin{figure}[!h]
	\centering
	\includegraphics[width=16cm]{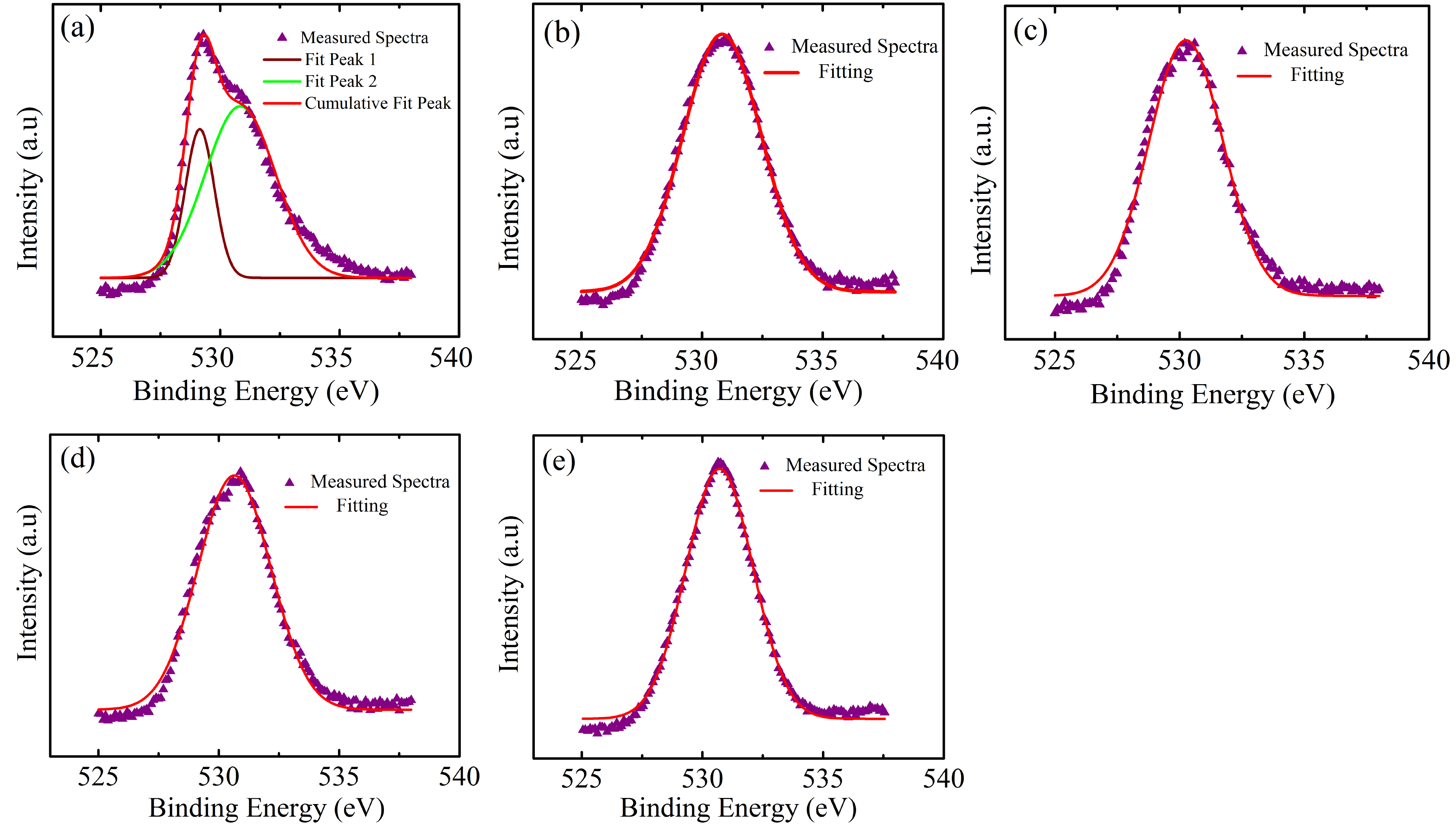}
	\caption{XPS spectra of the O 1s of Gd$_{0.7}$Sr$_{0.3}$MnO$_3$ (a) bulk polycrystalline powder materials, (b)  nanoparticles prepared by ultrasonication technique, (c-e) nanoparticles prepared by 4 h, 8 h and 12 h ball milling, respectively.} \label{fig3}
\end{figure}

Figure \ref{fig3} demonstrates the O 1s XPS spectra of (a) bulk polycrystalline  Gd$_{0.7}$Sr$_{0.3}$MnO$_3$  manganites, (b) ultrsonically prepared nanoparticles and (c-e) ball milled nanoparticles for 4 h, 8 h and 12 h, respectively. The O 1s XPS spectra of  bulk polycrystalline  Gd$_{0.7}$Sr$_{0.3}$MnO$_3$  manganites show a slightly asymmetric peak very close to 529.5 eV along with an additional peak. The asymmetric curves of the bulk sample can be Gaussian fitted by two symmetrical peaks at 529.3 ev and  530.8 eV, respectively (figure \ref{fig3} (a)). The lower binding energy peak at 529.3 eV corresponds to O 1s core spectrum, while higher binding energy peak is attributed to the oxygen vacancies i.e. to the oxygen related defects \cite{ref23,ref24} in the sample. Interestingly, in the case of Gd$_{0.7}$Sr$_{0.3}$MnO$_3$ nanoparticles prepared by ultrasonication we have observed a symmetrically single XPS peak (figure \ref{fig3} (b)) of O 1s \cite{ref21, ref25}. This indicates the absence of oxygen vacancy in ultrasonically prepared Gd$_{0.7}$Sr$_{0.3}$MnO$_3$ nanoparticles. Similar to the ultrasonically prepared nanoparticles, the ball milled nanoparticles do not show any peak corresponding to oxygen vacancy. It is expected that the presence and absence of oxygen vacancies \cite{ref93} in Gd$_{0.7}$Sr$_{0.3}$MnO$_3$ bulk materials and their nanoparticles, respectively will affect the mixed Mn$^{3+}$/Mn$^{4+}$ valence state which ultimately influences their magnetization \cite{ref94}. 

\subsection{Magnetic Properties}

\begin{figure}[!h]
	\centering
	\includegraphics[width=16 cm]{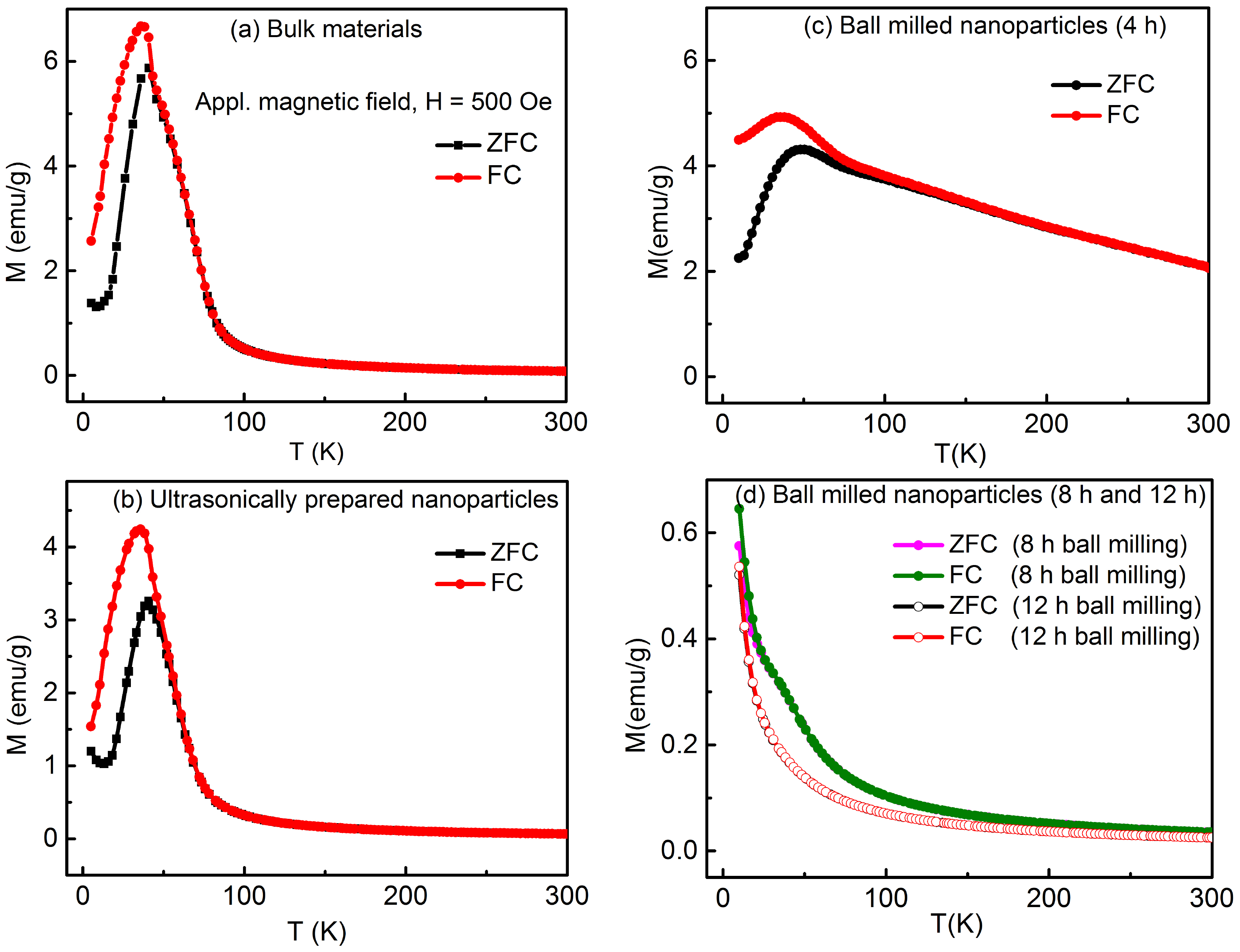}
	\caption {Temperature dependence of magnetization ($M-T$ curves) of Gd$_{0.7}$Sr$_{0.3}$MnO$_3$ (a) bulk materials, (b) ultrasonically prepared nanoparticles, (c) 4 h ball milled nanoparticles and (d) 8 h and 12 h ball milled nanoparticles in ZFC and FC processes.} \label{fig4}
\end{figure}

To investigate the  phase transition temperatures of bulk materials and corresponding synthesized nanoparticles, the temperature dependent magnetization measurements were carried out. Figures \ref{fig4} (a), (b) and (c) demonstrate the temperature dependent magnetization ($M-T$) curves of  Gd$_{0.7}$Sr$_{0.3}$MnO$_3$ bulk materials, and nanoparticles prepared by ultrasonication and by ball milling of bulk powder materials for 4 h, respectively in ZFC and FC processes. The temperature dependent ZFC and FC curves of 8 h and 12 h ball milled nanoparticles are shown in figure \ref{fig4} (d). To perform the experiment in the ZFC process, the sample was initially cooled from 300 K to 5 K and data were collected while heating in the presence of the applied field. On the other hand, in the FC mode, data were collected while cooling in the presence of a 500 Oe magnetic field. In the case of bulk (\textbf{figure \ref{fig4} (a)}) as well as nanoparticles prepared by ultrasonication technique (figure \ref{fig4} (b)), a clear splitting between ZFC and FC magnetization curves were observed below 42-47 K under the application of a magnetic field of 500 Oe. The splitting of ZFC and FC curves for 4 h ball milled nanoparticles started at higher temperature i.e. 70-75 K. The splitting of the curves suggests that Gd$_{0.7}$Sr$_{0.3}$MnO$_3$ bulk materials and their corresponding nanoparticles are in a spin-glass-like state \cite{ref52,ref54,ref57} at lower temperatures. Moreover, the temperature dependent magnetization curves exhibit a paramagnetic to spin-glass like splitting transition for both bulk powder materials and ultrasonically prepared nanoparticles at temperatures T$_g$ = 35-40 K. The value of the transition temperature is consistent with value reported elsewhere \cite{ref56} for the same manganite system. However, for 4 h balled milled nanoarticles, the phase transition is observed at a relatively higher temperature (38-48 K) as seen in figure \ref{fig4} (c). Moreover, on either side of this transition temperature, a different trend of the temperature variation of magnetization than that of bulk and ultrasonically prepared nanoparticles is observed. Notably, in the case of nanoparticles prepared by 8 h and 12 h ball milling, both ZFC and FC curves coincide with each other throughout the temperature range. Beside this, nanoparticles prepared by ball milling for milling time 8 h and 12 h does not show any phase transition, rather a small upturn to 0.6 emu/g is observed at low temperature. The steep increase of magnetization at low temperature may correspond to the paramagnetic contribution of the Gd$^{3+}$ moments \cite{ref426}. In this way, the magnetic behavior of the nanoparticles synthesized by ball milling technique is different from those of bulk materials and ultrasonically prepared nanoparticles.
\begin{figure}[!h]
	\centering
	\includegraphics[width=16 cm]{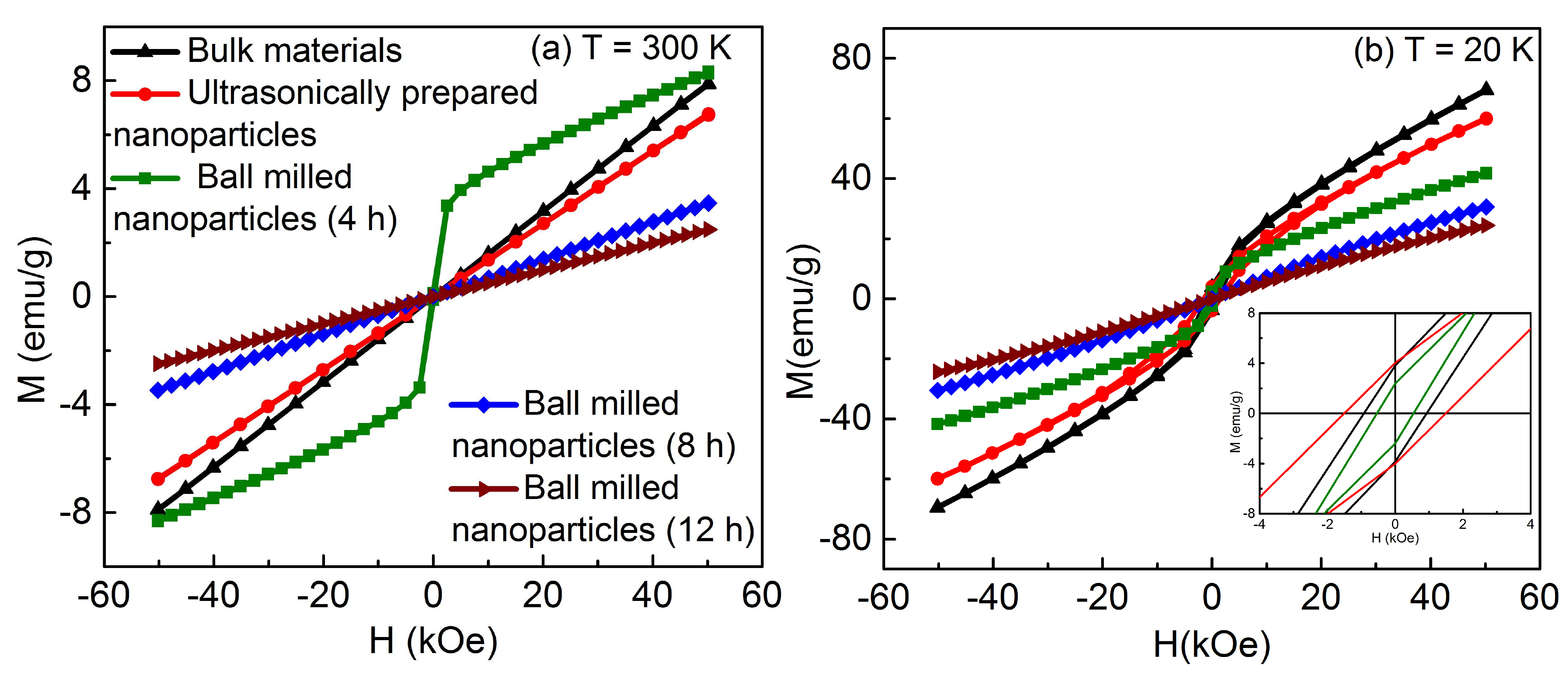}
	\caption{The $M-H$ curves of Gd$_{0.7}$Sr$_{0.3}$MnO$_3$ bulk materials, nanoparticles prepared by ultrasonication and ball milling (4 h, 8 h, 12 h) techniques at (a) 300 K and (b) 20 K. Inset of (b): an enlarged view of $M-H$ hysteresis loops showing the enhancement of the coercivity of the ultrsonically synthesized nanoparticles (red line).} \label{fig6}
\end{figure}

To further explore the difference between the magnetic properties of bulk materials and nanoparticles synthesized by two different techniques, we have carried out field dependent magnetization measurements. The magnetization vs magnetic field (M-H) hysteresis loops of Gd$_{0.7}$Sr$_{0.3}$MnO$_3$ bulk materials and nanoparticles prepared by ultrasonication and ball milling techniques were carried out at 300 K and 20 K, figures \ref{fig6} (a) and (b), respectively. At 300 K, the unsaturated linear curves without any detectable hysteresis as shown in figure \ref{fig6} (a) indicate the paramagnetic nature of Gd$_{0.7}$Sr$_{0.3}$MnO$_3$ bulk powder materials and their corresponding ultrasonically prepared and ball milled (8 h and 12 h) nanoparticles. However, for 4 h ball milled nanoparticles, a tiny hysteresis loop is observed at 300 K  which is clearly different from that of other materials under investigation. The maximum magnetizations $M_{s}$ of Gd$_{0.7}$Sr$_{0.3}$MnO$_3$ at 50 kOe for bulk materials and nanoparticles prepared by ultrasonication and ball milling are inserted in supplemental Table II \cite{ref99}. Notably, at 300 K, the magnetization of 4 h ball milled nanoparticles is anomalously higher compared to that of other materials and it is anticipated that the presence of 5.2 wt$\%$ Mn$_{2}$O$_3$ may influence the magnetic magnetic behavior of this sample. The Mn$_{2}$O$_3$ is an antiferromagnetic material with a N\'eel temperature of 80-100 K \cite{ref610}. We think that at 300 K, the 4h ball milling sample is dominated by superparamagnetic behavior, presumably due to Mn$_{2}$O$_3$. Apart from this anomalous behavior, the magnetization is higher in bulk materials compared to that of nanoparticles prepared by ultrasonication as well as ball milling for milling time 8 h and 12 h. The higher value of magnetization in bulk materials may be associated with its good crystallinity. In ultrasonically prepared nanoparticles, the amorphization is minimum compared to ball milled nanoparticles, and a reduction in magnetization is observed. For 8 h and 12 h ball milled nanoparticles, the amorphization is significant. Therefore, we think that the drastic decrease in magnetization in ball milled (8 h and 12 h) nanoparticles is associated with the amorphization of the ball milled nanoparticles. It is worth mentioning that similar class of nanoparticles of La$_{0.7}$Ca$_{0.3}$MnO$_3$ \cite{ref10} and La$_{0.8}$Sr$_{0.2}$MnO$_{3-\delta}$ \cite{ref11} manganites were prepared by using  ball milling from bulk polycrystalline samples. In these investigations, the particle size was found to decrease with milling time and for a longer milling time ($>$8 hours), the crystalline particle size was reduced. In addition to these, the ferromagnetic to paramagnetic transition temperatures, i,e, T$_c$ values of La$_{0.7}$Ca$_{0.3}$MnO$_3$ nanoparticles prepared by ball milling were found to decrease compared to those of unmilled bulk polycrystalline powder materials. Moreover, magnetization of the nanoparticles synthesized by ball milling was decreased dramatically compared to the bulk polycrystalline samples. A previous investigation also reported the magnetic properties of La$_{0.7}$Ca$_{0.3}$MnO$_3$ (LCM) thin films with structural disorder \cite{ref89}. In these reports, it was observed clearly that amorphous LCM thin film demonstrated a paramagnetic behavior whereas the crystalline film exhibited a cluster-spin-glass state \cite{ref89}. It is also reported that the properties of nanocrystalline manganites thin films are comparable with that of nanopowders \cite{ref81}.




The magnetization measurement at 20 K, figure \ref{fig6} (b), demonstrates hysteresis with significant coercivities in bulk materials and their corresponding nanoparticles prepared by ultrasonication technique. At 20 K, the 4 h ball milled nanoparticles also exhibits a hysteresis loop with a large coercivity. In this investigation, at 20 K, the coercive fields  ($H_c$) are 1500 Oe, 910 Oe and 540 Oe for ultrasonically prepared nanoparticles, and bulk materials and 4 h ball milled nanoparticles, respectively (inset of figure \ref{fig6} (b)). In fact, the coercive fields are temperature dependent \cite{ref26} and these increase with decreasing temperatures \cite{ref27}. Previous investigation also demonstrated that the coercive field increases with decreasing particle size \cite{ref326}. In our investigation, at 20 K, the coercive field is higher for ultrasonically prepared nanoparticles compared to that of bulk materials as was also observed in  \cite{ref61} for similar La$_{0.7}$Ca$_{0.3}$MnO$_3$ manganite system. In the case of nanoparticles prepared by ball milling technique with milling time 8 h and 12 h, the M-H curve is just a straight line with zero coercivity both at 300 K and 20 K as shown in figures \ref{fig6} (a) and (b), respectively.

In rare-earth based perovskite manganites, the presence of  Mn$^{4+}$ ions, due to the doping, enables itinerant e$_g$ electron of a Mn$^{3+}$ to hop to the neighbouring Mn$^{4+}$ through oxygen ion \cite{ref94} and thus mediate magnetism and conductivity. The presence of oxygen vacancies in Gd$_{0.7}$Sr$_{0.3}$MnO$_3$ bulk manganites is expected to enhance the conversion of a proportional number Mn$^{3+}$ to Mn$^{4+}$ \cite{ref93} and hence we observe higher value of the magnetization in bulk materials. Notably, the ultrasonically prepared nanoparticles are still good in terms of magnetization despite having apparently no oxygen vacancies. It may be expected from XPS measurement, figure \ref{fig3}, that from the surface of the nanoparticles, vacancies have been removed, while in the core of the nanoparticles it may still be present. More specifically, at 500 eV, the electron escape depth is about 5 nm and hence XPS indicates vacancies that have been removed from the particles' surface while the core volume may still have the vacancies of the bulk. 


\section{Conclusions} \label{II}
By using ultrasonication technique, rare-earth based  Gd$_{0.7}$Sr$_{0.3}$MnO$_3$ nanoparticles were produced directly from their bulk powder materials and the chemistry of the particles were likely unaltered apart from the absence of detectable surface oxygen vacancies. Hence, bulk and nanoparticle materials were more directly comparable in terms of  magnetic properties than materials produced by two different synthesis techniques \cite{ref28}. The nanoparticles were also synthesized by ball milling of bulk powder materials prepared by solid state reaction technique. The weight \% of the major orthorombic (pbnm) phase of ultrasonically prepared nanoparticles was comparable to that of bulk materials.  For ball milled nanoparticles, the crystalline to amorphous phase conversion gradually increased with milling time. The transition temperature (35-40 K) of ultrsonically synthesized nanoparticles and corresponding bulk materials is unaltered. The magnetization of the nanoparticles prepared by 4 h ball milling is anomalously different as compared to that of other materials. The magnetization of the nanoparticles prepared by 8 h and 12 h ball milling is found to decrease drastically compared to the ultrasonically prepared nanoparticles as well as their bulk materials which may be associated with the amorphization and contamination of the ball milled nanoparticles.  Some amorphization is also observed in ultrasonically prepared nanoparticles. Nevertheless, the size, crystallinity, amount of major phase formation, unaltered phase transition etc of ultrasonically prepared Gd$_{0.7}$Sr$_{0.3}$MnO$_3$ nanoparticles demonstrate the potential of this synthesis route for the fabrication of these complex perovskites.  To evaluate the potential of rare-earth based novel manganite nanostructures for future technical applications, the knowledge of their synthesis and the correlation of their structural and magnetic properties are essential.  We believe that the demonstrated ultrasonication technique may be promising for fabrication of other rare-earth based manganite nanoparticles and can be further developed as a versatile technique for the preparation of nanoparticles of a wide range of materials. 




\section{Acknowledgements}
This work was supported by Ministry of Science and Technology, Government of Bangladesh, Order No: 39.009.002.01.00.053.2014-2015/PHY'S-273/. The Institute for Molecular Science (IMS), Japan is sincerely acknowledged for providing facilities of SQUID magnetometer and Professor F. A. Khan for providing planetary ball milling facility.

\end{document}